\documentclass[12pt]{JHEP3}

\usepackage{amsfonts, latexsym}
\usepackage{amsmath, amsthm}
\usepackage{graphicx}
\usepackage{epstopdf}

\newcommand{\be}{\begin{equation}}
\newcommand{\ee}{\end{equation}}
\newcommand{\bea}{\begin{eqnarray}}
\newcommand{\eea}{\end{eqnarray}}
\newcommand{\ds}{\displaystyle}

\def\d{\partial}
\def\half{\frac{1}{2}}
\newcommand{\bel}[1]{\begin{equation}\label{#1}}
\newcommand{\beal}[1]{\begin{eqnarray}\label{#1}}

\newcommand{\rf}[1]{(\ref{#1})}

\def\mps{M_P^2}

\title{Inflationary predictions at small $\gamma$}

\author{Ewa Czuchry \footnote{Email: eczuchry@fuw.edu.pl} \\
So\l tan Institute for Nuclear Studies\\
ul. Ho\.za 69,
00-681 Warszawa, Polska.
}

\abstract{
This paper explores single field inflation models with a constant, but
arbitrary speed of sound $c_s$, obtained by deforming the kinetic energy terms
to a Dirac-Born-Infeld form. Allowing $c_s<1$ provides a simple
parameterization of non-gaussianity. The dependence of inflationary observables on
the parameter $c_s$ is considered in the leading order slow roll
approximation. The results show that in most cases the dependence is
actually rather weak for the range of $c_s$ allowed by existing bounds on
non-gaussianity. 
}

\preprint{}

\begin{document}

\section{Introduction}

Models of k-inflation \cite{ArmendarizPicon:1999rj,Garriga:1999vw} have
been the subject of much recent 
interest. These
models modify standard single-field inflation by allowing 
a more general form of the kinetic energy terms.
An important feature of k-inflation is the alteration of the speed
of propagation of disturbances in the inflaton field -- the speed of
sound  $c_s$. Models of inflation with canonical kinetic energy have
$c_s=1$, but with a more general form of kinetic energy this is no longer
the case -- the speed of sound i a field dependent quantity.

One particularly interesting example of this is  a model which replaces
the canonical kinetic energy by the Dirac-Born-Infeld form
\cite{Silverstein:2003hf,Alishahiha:2004eh}, which has come 
to the fore in recent years in the context of D-brane inflation
\cite{Cline:2006hu}--\cite{McAllister:2007bg}.
It is however also quite natural to view models of this kind in the general
context of k-inflation \cite{Lidsey:2006ia,Spalinski:2007qy}. 
The DBI form of the kinetic energy terms
involves a square root factor, $\gamma>1$, reminiscent of the Lorentz
factor of special relativity. Indeed, the square root is responsible for
introducing a ``speed limit'' on the inflaton scalar. This facilitates a
form of non-slow-roll inflation \cite{Alishahiha:2004eh}, on which most
effort has focused. For such 
trajectories of the inflaton the Lorentz factor 
$\gamma$ is large. However, it is also interesting to consider what
quantitative effect allowing $\gamma>1$ has in the slow roll regime. In the
case of DBI models the speed of sound is expressed in terms of $\gamma$ as
$c_s=1/\gamma$.

A particularly simple situation arises when the $\gamma$ factor is constant
\cite{Spalinski:2007un}, which means that the speed of sound is also
constant, as in the 
canonical case, but no longer equal to the speed of light. This case can be
considered as a leading approximation in an expansion of the field
dependent speed of sound if it is assumed to vary slowly in the relevant
region of field space. Some exactly solvable examples of this sort have
recently been discussed in \cite{Spalinski:2007un}--\cite{Lorenz:2008et}. 

From a phenomenological perspective the interest in k-inflationary
models stems from the fact that they provide a fairly simple way
of accounting for non-gaussianity \cite{Bartolo:2004if} in the spectrum of
density 
perturbations. So far no conclusive evidence for departures from
gaussianity has been observed, but it is of great interest to see if, and what kind of,
 theoretical possibilities exist, given the high expectations
that data which will become available in the coming years will
allow for a significant tightening of existing bounds. The
simplest measure of non-gaussianity is the parameter $f_{NL}$,
which is defined in terms of 3-point functions in
\cite{Chen:2006nt,Creminelli:2006rz} (for 
example). For DBI models one has the simple relation \cite{Chen:2006nt}
$f_{NL}=35 (\gamma^2 - 1)/108$, so the value of $\gamma$ strongly affects the
deviations from gaussianity. The current bounds on
$f_{NL}$ imply roughly $\gamma<35$. 
DBI models with constant speed of sound provide a
very simple parameterization of non-gaussianity in models of single
field inflation.
Even if the observed non-gaussianity
turns out 
to be large, the variation of the speed of sound with scale could
be negligible.

This note reports the results of analyzing a series of frequently
considered models of slow roll inflation and explores how
sensitive their predictions are when one allows a small deviation
from the canonical form of the kinetic energy, as measured by a
constant $\gamma>1$. The observables $n_s$ and $r$ 
have been computed for general models of k-inflation by Garriga and
Mukhanov
\cite{Garriga:1999vw}, and they are easily adapted to the case of DBI
kinetic energy \cite{Chen:2006nt}. These formulae are 
evaluated at the time  when the present Hubble scale crossed the horizon during
inflation. The corresponding number of e-folds is denoted by
$N_\star$\footnote{As in a number of 
other studies, the number of e-folds is
  defined as decreasing to 0 at the end of inflation.}. This number is
afflicted by some theoretical uncertainties \cite{Kinney:2005in}, and the
possibility of having  
$\gamma>1$ adds another ones, as discussed in section
\rf{basics}. The remaining sections discuss a number of popular
inflationary models case by case. In most cases, notably chaotic inflation,
the results for the inflationary observables do not depend on $\gamma$, or
the dependence is very weak. However in some cases of modular inflation it
is found that the tensor fraction $r$ effectively grows with $\gamma$, so
one can envisage that it might become observable due to this effect.

\section{Basic formalism}
\label{basics}

The simplest models assume that inflation is driven by a single scalar
field, whose contribution to the energy density dominates and leads to the
negative pressure which drives the accelerated
expansion. Under very general assumptions
\cite{ArmendarizPicon:1999rj,Chen:2006nt} the inflaton action 
takes the form
\be
S = \half \int \sqrt{-g} \big(R + P(X,\phi)\big)\ ,
\ee
where $X=\half(\d\phi)^2$.

Brane inflation models in string theory
\cite{Cline:2006hu}-\cite{McAllister:2007bg} 
have attracted much 
attention to a specific example, the
Dirac-Born-Infeld action:
\bel{dbi}
S = - \int d^4x\ a(t)^3\ \{f(\phi)^{-1}(\sqrt{1-f(\phi) \dot{\phi}^2}-1) +
V(\phi)\}  \  .
\ee
The function $f$ appearing here can be related to the compactification
geometry of the D-brane model.
The action \rf{dbi} leads to field equations
for a perfect fluid 
\bea
\dot{\rho} &=& - 3 H (p+\rho)\ , \label{conserv}\\
3\mps H^2&=& \rho \label{friedman} \ ,
\eea
with
\bea
p     &=& \frac{\gamma-1}{f\gamma} - V(\phi)\ ,\\
\rho  &=& \frac{\gamma-1}{f} + V(\phi) \ , 
\eea 
where 
\be \gamma =
\frac{1}{\sqrt{1-f(\phi)\dot{\phi}^2}} \ . 
\ee 
It is convenient to
express these equations in the Hamilton-Jacobi form
\cite{Kinney:1997ne,Spalinski:2007kt} 
\bel{hjdbi} 
3\mps H^2 - V =  \frac{\gamma-1}{f} \ . \ee 
where now
\bel{gamma} 
\gamma(\phi)=\sqrt{1+ 4 M_P^4 f H'^2} \ . 
\ee 
To
quantify the conditions under which inflation takes place one
defines
\bel{epsdef} 
\varepsilon = \frac{2\mps}{\gamma}
\left(\frac{H'}{H}\right)^2 \ . 
\ee
The condition for the Universe
to be inflating is $\varepsilon<1$. Just as in the canonical case,
the leading order slow roll approximation entails dropping the
$H'$ dependence in \rf{hjdbi}, which means taking \bel{losr} 3\mps
H^2 = V\ . \ee It is straightforward to write down corrections to
observables related to the primordial perturbation spectra given
the results of Garriga and Mukhanov \cite{Garriga:1999vw}: the
spectral indices can be written as follows:
\beal{indices}
n_s-1&=&-2\varepsilon + \eta + \sigma\  ,\\
n_T &=& -2 \varepsilon \ .
\eea
where
\bea
\eta &=& \frac{4\mps}{\gamma}\frac{H''}{H} - 2\varepsilon + \sigma\ ,\\
\sigma &=& -\frac{2\mps}{\gamma}\frac{H'}{H}\frac{\gamma'}{\gamma} \ .
\eea
In the above formulae the right hand side is to be evaluated at horizon
crossing, i.e. at $k=aH\gamma$. In the slow roll approximation one has
\bea
\varepsilon &=& \varepsilon_V \ ,\\
\eta &=& 2\eta_V - 4\varepsilon_V\ ,
\eea
where
\bea
\varepsilon_V &=& \frac{\mps}{2\gamma}\left(\frac{V'}{V}\right)^2\ , \\
\eta_V &=& \frac{\mps}{\gamma} \frac{V''}{V}\ ,
\eea
are the potential slow roll parameters. These have the usual dependence on
the potential, but include a factor of $1/\gamma=1$. For the case of
constant $\gamma$ one also has 
$\sigma=0$ in \rf{indices}, so that 
\beal{slowindices}
n_s-1&=&-6\varepsilon_V + 2 \eta_V\ , \\
n_T &=& -2 \varepsilon_V \ .
\eea
The tensor to scalar ratio is \cite{Chen:2006nt} 
\be
r = \frac{16\varepsilon_V}{\gamma} \ .
\ee
One way to proceed is to express everything as a function of $N$ --
the number of e-folds. Using the convention that $N$ decreases
during inflation, reaching $N=0$ at the end of inflation, one has 
$dN=-Hdt$, which leads to the formula 
\bel{nform} 
N(\phi) =
-\frac{1}{\sqrt{2}} \int_{\phi}^{\phi_e} \frac{d\phi}{M_P}
\sqrt{\frac{\gamma}{\varepsilon}} \ .
\ee 
The value of $N$ at
horizon-crossing (denoted by $N_\star$) can be determined as
explained, for example, in \cite{Dodelson:2003vq,Liddle:2003as}. The
essence is that 
$N_\star$ is given by 
\be
N_\star = 50 + \delta N +
\log\left(\frac{H_{end}}{H_\star}\right) + \log(\gamma) \ .
\ee 
where
$\delta N = \pm 20$ accounts for various
uncertainties in the post-inflationary evolution of the Universe
\cite{Alabidi:2005qi} (such as the reheating temperature). The last two
terms reflect 
features of inflation itself. The last term is a consequence of
the change of speed of sound -- horizon crossing is at
$k=aH\gamma$. The next to last term accounts for the decrease in
energy density during inflation. It was discussed in
\cite{Dodelson:2003vq,Liddle:2003as}. 
It is easy to show that in the DBI case one has
\bel{logint}
\log\left(\frac{H_{end}}{H_\star}\right) = \int_0^{N_\star} dN\ 
\varepsilon(N) \ . 
\ee
This has the same form as in the case of
canonical kinetic energy \cite{Dodelson:2003vq} (but of course the
$\varepsilon$ appearing 
here contains the $\gamma$ factor as in \rf{epsdef}). While this
term is zero for de Sitter expansion, it could in principle give a
non-negligible contribution to \rf{nstar}. In such a case
\rf{nstar} is an equation which has to be solved for $N_\star$. In
the examples considered in this paper the resulting shift of $N$
is usually not large in the allowed range of $\gamma$: is can be
of order of a few at most, so it is smaller than the uncertainties
in $\delta N$. It does however affect the window of reasonable
values of $N_\star$ which is usually considered
\cite{Alabidi:2005qi,Kinney:2005in}. 

In the sequel we will shift the number at which we evaluate $N_{\star}$  closer to the upper limit of estimated values of e-folds' number, i.e. we will use the formula:
\bel{nstar} 
N_\star = 60 +\log\left(\frac{H_{end}}{H_\star}\right) + \log(\gamma)\ ,
\ee
where $\delta N$ has been partially absorbed in a number 60.

\section{Large field models}

\subsection{Chaotic inflation}

Chaotic inflation \cite{Linde:1983gd}
has become one of the most important examples of
inflation. This is due partly to its simplicity -- essentially any
potential will work, and for simple monomial examples there is
only one parameter, which can be fixed using the COBE
normalization condition. For a quadratic monomial potential the model then
predicts the scalar index, 
and the prediction is consistent with most recent data
\cite{WMAP5}. The simplicity of chaotic inflation is not to be
held against it: one can envisage it arising as an effective
description in more complex contexts, such as supergravity or string theory
\cite{McAllister:2008hb}. The fact that chaotic
inflation with a monomial potential gives a firm prediction for
the scalar index makes it particularly interesting from the point
of view of this note: it is natural to ask how sensitive this
prediction is to deformations of the kinetic term.

The relevant potential reads: 
\be
V=\Lambda\left(\frac{\phi}{\mu}\right)^p\ , 
\ee 
In the slow roll
approximation one has: 
\be
\varepsilon_V=\frac{M_P^2}{2\gamma}\left(\frac{V'}{V}\right)^2
\approx \frac{k^{2}}{2\gamma} \frac{M_P^2}{\phi^{2}} \ .\ee We need
to  express $\varepsilon_H$ in terms of $N$ and evaluate it at
horizon-crossing. $N$ is given in terms of $\phi$ by the formula
\rf{nform} 
\be N  =
\frac{\gamma}{2k}\left(\frac{\phi^{2}}{M_{P}^{2}} -
\frac{\phi^{2}_{e}}{M_{P}^{2}} \right) \ .
\ee 
Inflation ends when
$\varepsilon_{H}=1$, i.e. when the inflaton field reaches the
value $\frac{\phi^{2}_{e}}{M_{P}^{2}}=\frac{k^{2}}{2\gamma}$.
Hence: 
\be
\frac{\phi^{2}}{M_{P}^{2}}=\frac{k}{\gamma}\left(2N+\frac{k}{2}\right)\ .
\ee This determines $\varepsilon_{V}(N)$ as \be
\varepsilon_{V}=\frac{k}{4N+k}\ . 
\ee 
The dependence on $\gamma$
drops out in the above formula, so the scalar index does not
explicitly depend on the value of $\gamma$. The fact is that
$\gamma>1$ only enters in by shifting the window of allowed values
of ${N_{\star}}$:
\be 
\log(H_i/H_e) = - \int_0^{N_{\star}} dN \varepsilon_V(N)\ . 
\ee
This shift is quite small and has negligible effect on the observable
quantities. 
Thus, at this level of approximation the predictions of chaotic
inflation remain unchanged.

\subsection{Natural inflation}

Natural Inflation \cite{natural} is a model which arises from the
assumption that the inflaton is an axion-like field, whose potential is
generated by instanton effects:
\be
V=\frac12V_0\left(1-\cos\left(\sqrt{2|\eta_0|}\frac{\phi}{M_p}\right)\right)\
.\ee 
Such a
potential appears provides an example of a theoretically-motivated
mechanism for generating chaotic inflation \cite{Boubekeur:2005zm}, since
chaotic 
inflation in the case $k=2$ may be considered as an 
approximation to a harmonic potential near its maximum. 

The slow roll parameters read: 
\begin{align}
\varepsilon_V&=\frac{|\eta_0|}{\gamma}\cot^2
\left(\sqrt{\frac{|\eta_0|}{2}}\frac{\phi}{M_p}\right)\ ,\\
\eta_V&=\varepsilon_V-\frac{|\eta_0|}{\gamma}\ .
\end{align}
To compute the inflationary observables one needs to assess the
impact of $\gamma>1$ on the allowed range of $N_\star$. Using
\rf{logint} one finds 
\be 
\ln H_i/H_e =
-\frac{|\eta_0|}{\gamma}{N_{\star}}+\frac12\ln\frac{\ds
\frac{|\eta_0|+\gamma}{\gamma}
\exp\left(2\frac{|\eta_0|}{\gamma}N_{\star}\right)-1  }{\ds
\frac{|\eta_0|+\gamma}{\gamma}-1}\ , 
\ee 
which gives a shift of ${N_{\star}}$
ranging from 62 (for $\gamma=1$) to 66 (for $\gamma=35$). This
shift of $N_\star$ lays well within the uncertainty due to unknown
features of post-inflationary evolution (such as details of the
reheating stage). Therefore, for the remaining calculations of
natural inflation, the dependence of $N_\star$ will be suppressed
with the understanding that the allowed range is shifted by this amount.  

Although $N_\star$ depends weakly on $\gamma$, the spectral index
$n_{s}-1$, which is a physical observable (in contrast to $N_\star$) is 
a function of both $\gamma$ and ${N_{\star}}$: 
\be
n_{s}-1=-6\varepsilon_V+2\eta_V=-\frac{\ds
\frac{4|\eta_0|}{\gamma}}{\ds \frac{|\eta_0|+\gamma}{\gamma}
\exp\left(2\frac{|\eta_0|}{\gamma}{N_{\star}}\right)-1}
-2\frac{|\eta_0|}{\gamma}\ .\label{ns-nat} 
\ee 
This relation brings
more significant change on the predicted by  theory values of
$n_{s}-1$ -- around 10\% for $\gamma=35$. The value of parameter
$|\eta_0|$ was taken such as to mimic WMAP5 results \cite{WMAP5}
for $\gamma=1$, i.e.  $-\eta_{0}=0.014$.  In this model, $n_s(\gamma)$ is an increasing
function and tends to 1. However, the spectral tilt depends
effectively on the ratio $|\eta_0|/\gamma$, so 
unless one has an independent estimate of $|\eta_0|$ in a specific model
one can just fix this ratio using the observed value of $n_{s}-1$. 

For the tensor fraction $r$ one has
\be 
r=\frac{16\varepsilon_V}{\gamma}=16\frac{\ds
\frac{|\eta_0|}{\gamma^2}}{\ds \frac{|\eta_0|+\gamma}{\gamma}
\exp\left(2\frac{|\eta_0|}{\gamma}{N_{\star}}\right)-1} \ .
\ee 
This depends on $\gamma$ and $|\eta_0|/\gamma$ separately. If the value of 
$|\eta_0|/\gamma$ if fixed using the measured value of $n_s-1$,
then $r(\gamma)\sim 1/\gamma)$. Taking into account the weak dependence of 
$N_\star$ on $\gamma$ (but with fixed
$|\eta_0|/\gamma$) does not significantly alter the conclusions.

\section{Small field models}

\subsection{New and modular inflation}

In models of this type the inflaton field is usually assumed to be moving
away from $\phi=0$ and for some time the potential can be approximated by
a polynomial of the form \cite{Alabidi:2005qi, mod1, mod2}:
\be
V=V_{0}\left(1-\frac{\phi^k}{\mu^k}\right) \ .\label{new-pot}
\ee
This could be, for example, the Taylor expansion of potential arising from
a phase transition associated with spontaneous symmetry breaking.  
This expansion is taken near unstable equilibrium at the origin, with
$k$ being the lowest non vanishing derivative there. It is assumed that
$\phi\ll M_P$, and that the constant term dominates the potential. 

To assess the impact of a having a lower speed of sound it is convenient to
consider  the cases $k=2$ and $k\ge 3$ separately.

\subsubsection{$k=2$}

In this case it is convenient to rewrite the inflation potential
\eqref{new-pot} in the form 
\be
 V=V_{0}\left(1-\frac12|\eta_0|\frac{\phi^{2}}{M^2_p}\right)\ ,
\ee
where $\eta_{0}<0$ is the value of the parameter $\eta_V$ at the 
maximum of the potential. Thus, in the slow roll approximation 
\be
\phi=\phi_e\exp\left(-\frac{|\eta_0|}{\gamma}N \right)\ ,
\ee
and
\bea
\varepsilon_V &=& \frac{|\eta_0|^2}{2\gamma}
\frac{\phi_e^2}{M_P^2}\exp\left(-\frac{2|\eta_0|}{\gamma}N 
\right)\ ,\\
\eta_V &=& \frac{\eta_0}{\gamma}\ .
\eea
Here the value of the field $\phi_{e}$ at the end of inflation is kept as a
parameter. This is 
due to the specific property of the scalar potential with a leading
quadratic term \cite{Dodelson:1997hr,Boubekeur:2005zm}.   
Namely, the small field assumption $\phi\ll M_{P}$, valid when the
observable perturbations are generated, is not fulfilled during the whole
inflation phase. 
The condition $\varepsilon=1$ which is usually used to determine $\phi_{e}$  
would in this case imply $\phi_{e}\gg M_{P}$. One expects however that the
non-leading terms in the potential start playing an important role long
before. Therefore, following \cite{Dodelson:1997hr,Boubekeur:2005zm}, we
shall assume that due to these contributions the potential 
steepens and inflation ends when the inflation field
$\phi$ is of order of the Planck's mass $M_{P}$. 

Using the same procedure as in the
previous section, the contribution to $N_\star$ in \rf{nstar} is 
\be
\ln H_i/H_e= \frac{|\eta_0|}{4}\frac{\phi_e^2}{M_P^2}\left[ \exp
  \left(-\frac{2|\eta_0|}{\gamma}{N_{\star}}
\right)-1\right]\ .
\ee
Since the spectral tilt $n_{s}-1$ does not depend directly on $N_\star$:  
 \be
n_s-1\approx2\eta_H=-2\frac{|\eta_0|}{\gamma}\ .
\ee 
(where we have taken into account $\varepsilon_V\ll\eta_V$), one can
determine the ratio $\eta_0/{\gamma}$ using the WMAP5 result for $n_s-1$. 
One finds $\eta_0/{\gamma}=0.02$. It can then be seen that the shift of
$N_\star$ is then no more than $4$ efolds (allowing $1\le\gamma\le35$).   

The tensor fraction $r$ reads  
\be
r = 8\frac{|\eta_0|^2}\gamma^2\frac{\phi_e^2}{M_P^2}
\exp\left(-\frac{2|\eta_0|}{\gamma}{N_{\star}}\right) =   
2\frac{(n_{s}-1)^2}\gamma\frac{\phi_e^2}{M_P^2}\exp\left((n_{s}-1){N_{\star}}\right)\ ,
\ee 
and for $\phi_e<M_P$ is unlikely to be observable. The small 
variation of ${N_{\star}}$ with $\gamma$ does not alter this situation.

\subsubsection{$k=3$ and higher}

Here the results strongly depend on the value of the scale 
$\mu$. For $\mu\ll M_{p}$ the spectral tilt does not depended 
on $\gamma$: one finds 
\be \label{spec-3} n_{s}-1=-\frac{k-1}{k-2}\frac2N_\star\ ,
\ee 
which is the standard result \cite{Boubekeur:2005zm}. 

The shift of $N_\star$ can be calculated as before and the result is 
\be
\ln H_i/H_e = \frac12\left(\frac{k}{\sqrt{2\gamma}}\frac{M_p}{\mu}
\right)^{\frac{k}{1-k}} \left\{\left[(k-2)2^{\frac{k-2}{2k-2}}
  \frac{k^{\frac{1}{k-1}}}{\gamma^{\frac{k}{2k-2}}} 
\left(\frac{\mu}{M_p}\right)^{\frac{k}{1-k}} {N_{\star}}+ 1
\right]^{\frac{k}{2-k}}-1\right\} \ .
\ee 
This and the $\log \gamma$ term, give altogether $N_\star$ changing with  
$\gamma\in[1,35]$, for $k=3$ from 60 to 65. This range gets narrower  with decreasing
 ratio $\mu/M_P$. However, as $n_s-1\sim
1/N_\star$, those modifications on $N_\star$ do not have strong impact on
vales of spectral tilt.

If one allows $\mu$ being close to the Planck's mass both the  spectral tilt and
tensor fraction become quite sensitive to the value of 
$\gamma$. One finds 
 \be
n_{s}-1=\frac{-2(k-1)}{(k-2){N_{\star}}+k^{-\frac1{k-1}} \gamma^{\frac{k}{2k-2}}
  \left(\frac{\mu}{M_{p}} \right)^{\frac{k}{k-1}} 
} \  ,
\ee
and 
\be
r=\frac{16}\gamma\left[(k-2)2^{\frac{k-2}{2k-2}}{k^{\frac{1}{k-1}}}
  {\gamma^{\frac{k}{2-2k}}} 
  \left(\frac{\mu}{M_p}\right)^{\frac{k}{1-k}} {N_{\star}}+ 1
  \right]^{\frac{2k-2}{2-k}}\ , 
\ee 
Increasing $\gamma$ raises the value of
$n_{s}$ by a few percent and moves it closer to 1. For example, for $\mu\sim
M_p$, $k=3$ and ${N_{\star}}=60$, $n_s-1$ varies from -0.066 ($\gamma=1$) to -0.058
($\gamma=35$). 
The tensor fraction $r$, for $k=3$, $\gamma=1$,
$\mu\sim M_P$ and ${N_{\star}}=60$ is of the order $10^{-7}$, and even
smaller for  $\mu\approx 0.1 M_P$ -- around $10^{-13}$. However, the effect of 
increasing $\gamma$ is quite significant, raising $r$ to around
$10^{-5}$ (when $\mu\sim M_P$), which is perhaps only an order of magnitude  
away from being observable \cite{r-obs}. 
This effect gets weaker for higher $k$, as well as for 
smaller $\mu/M_P$. 

The plot below shows $r(\gamma)$ for $k=3$, ${N_{\star}}=60$,  $\mu\sim M_P$.

\begin{FIGURE}
\centering
\includegraphics[height=70mm]{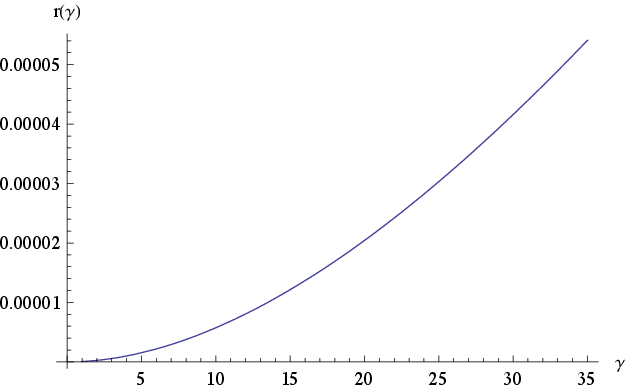}
\end{FIGURE}

\subsection{$F-$ and $D-$term inflation}

Another type of inflaton potential which is frequently considered 
appears in the context of supersymmetric models of inflation. 
For
the simplest choice of K\"ahler potential, 
either $F-$term \cite{f-term} or $D-$term
\cite{stewart} inflation is possible. 
The leading
term in the potential is constant, and the dependence on the inflaton field
arises as a loop correction. 
In many cases the relevant potential takes the form:
\be
V(\phi)=V_0\left(1+\frac{\lambda^2}{8\pi^2}\ln\frac{\phi}{Q}\right)\ ,
\ee 
where $\lambda$ is a coupling to the waterfall field and the 
renormalization scale $Q$ is of the order of the inflationary
field $\phi\gg\phi_{e}$. 

In this case the predictions depend on $\gamma$ very weakly. The number of
e-folds ${N_{\star}}$  does not 
depend on a Lorentz factor up to order $\lambda^{2}$.
The tensor fraction is a decreasing function of $\gamma$: 
\be r=\frac{\lambda^{2}}{2\pi^{2}\gamma}\frac 1{N_{\star}}\approx0.0011\,
\frac{\lambda^{2}}{\gamma}\left(\frac{50}{{N_{\star}}}\right)^{2} \ ,
\ee
and for small $\lambda$ much bellow 1, and bigger values of $\gamma$,
values of $r$ fall into unobservable region.

\subsection{Exponential potential}

The final model considered here is the exponential potential 
\be
V(\phi)=V_0\left(1-\exp\left(-q\frac{\phi}{M_p}\right)\right) \ ,
\ee
which appears in the context of 
supergravity models (with the value $q=\sqrt{2}$) by choosing the K\"ahler
potential appropriately (\cite{stewart}). It may also arise (with 
$q=\sqrt{2/3}$) in models of inflation driven by non-Einstein gravity 
\cite{non-einstein}, although then, strictly speaking, one is on the 
border of the small-field regime.

In case of this  potential, the $\gamma$-dependence of $N_\star$ is
negligible for either
value of $q$, since  
\be 
-\ln
H_i/H_e = \frac{1/2}{\ds\frac{q^2}{\gamma}{N_{\star}} +
  \frac{q+\sqrt{2\gamma}}{\sqrt{2\gamma}}} \approx 
\frac{\gamma}{2q^{2}{N_{\star}}} < 1 \ .
\ee
However, the spectral index $n_{s}-1$: 
\be n_{s}-1\approx
2\eta_{H}=-\frac{2}{\ds
{N_{\star}}+\frac{\gamma}{q^{2}}+\sqrt{\frac{\gamma}{2q^{2}}}} \ ,
\ee 
shows rather insignificant dependence on $\gamma$: in supersymmetric model
($q=\sqrt{2}$), for a fixed ${N_{\star}}=60$, $n_{s}-1$ increases from
$-0.033$ ($\gamma=1$) to $-0.032$ for $\gamma=35$. The situation is similar
for the case of non-Einstein gravity inflation.

However, the tensor fraction $r$:
\be 
r(\gamma,{N_{\star}})=
 \frac{\ds 8}{\ds q^2\left({N_{\star}}+\frac{\gamma}
   {q^{2}}+\sqrt{\frac{\gamma}{2q^{2}}}\right)^2} \ ,
\ee
is very sensitive to changing
$\gamma$, and although $r(\gamma)$ is a decreasing function, it
mainly  remains in a range that is likely observable, i.e. it
varies from 0.001 to 0.0006 with $N_\star$ fixed at
60. Smaller values of $N$ slightly increase the estimate of $r$.

\section{Conclusions}

It is expected that soon the existing bounds on non-gaussianity will be
significantly tightened or a measurement of it will be made. In view of
this it is important to consider theoretical options which lead to
non-gaussian perturbation spectra. One simple possibility was considered
here. The conclusion is basically that as long as the speed of sound is
constant, the basic predictions for the spectral tilt and tensor fraction
are quite robust even if one allows $f_{NL}$ as large as existing limits
permit. Some variation of the tensor fraction with $f_{NL}$ is possible 
however, and while generically $r$ goes down with rising non-gaussianity,
in some examples (new inflation with $k>2$) it can increase somewhat. 

\begin{center}
{\bf Acknowledgements}
\end{center}

\noindent I would like to thank Micha\l\ Spali\'nski for helpful
discussions  and pointing out the problem. This work has been supported by a grant PBZ/MNiSW/07/2006/37.

\end{document}